%% file: GLOBECOM25.tex
\documentclass[conference]{IEEEtran}

\IEEEoverridecommandlockouts
\input{header}
\begin{document}
\input{title-contents.tex}
\maketitle
\input{abstract.tex}
\section{Introduction}
\label{sec:intr}
\input{section1}

\section{System Model}
\label{sec:syst_mode}
\input{section2}
\section{Channel Modeling}
\label{sec:chan_mode}
\input{section3}
\subsection{PDE Formulation for Molecular Concentration}
\label{subsec:form}
\input{section3a}
\subsection{Approximate PDE Solution in Flow-Dominated Regime}
\label{subsec:appr}
\input{section3b}
\subsection{Arrival Probability and Arrival Rate Modeling}
\label{subsec:deri}
\input{section3c}
\section{Results and Analysis}
\label{sec:resu}
\input{section4}
\section{Conclusion}
\label{sec:conc}
\input{section5}
\section*{Acknowledgment}
This work is supported by the Korean government (RS-2023-00208922, 2022R1A5A1027646).
\bibliographystyle{IEEEtran}
\bibliography{IEEEabrv,GlobeCom25}
\end{document}

%% file: header.tex
\usepackage[letterpaper,top=0.7in,bottom=1.05in,left=0.625in,right=0.625in]{geometry}
\usepackage{cite}
\usepackage{amsmath,amssymb,amsfonts,multirow}
\usepackage{algorithmic}
\usepackage[pdftex]{graphicx}
\usepackage{caption}
\usepackage{subcaption}
\usepackage{textcomp}
\usepackage{xcolor}
\usepackage{url}
\usepackage{array}
\usepackage[caption=false,font=footnotesize]{subfig}
\usepackage{siunitx}
\usepackage[normalem]{ulem}

\usepackage{comment}

\def\BibTeX{{\rm B\kern-.05em{\sc i\kern-.025em b}\kern-.08em
		T\kern-.1667em\lower.7ex\hbox{E}\kern-.125emX}}

\newcommand{\tc}{\textcolor}

%% file: title-contents.tex
\title{
Three-Dimensional Channel Modeling for Molecular Communications in Tubular Environments with Heterogeneous Boundary Conditions
}
\author{
    \IEEEauthorblockN{
        Yun-Feng Lo\IEEEauthorrefmark{1}\thanks{Y.-F. Lo and C. Lee contributed equally to this work.}, 
        Changmin Lee\IEEEauthorrefmark{2}, and Chan-Byoung Chae\IEEEauthorrefmark{4}
    } 
    \IEEEauthorblockA{\IEEEauthorrefmark{1}Georgia Institute of Technology, Atlanta, GA, 30332 USA}
    \IEEEauthorblockA{\IEEEauthorrefmark{2}Newratek, Seoul, 06175 South Korea}  
    \IEEEauthorblockA{\IEEEauthorrefmark{4}School of Integrated Technology, Yonsei University, Seoul, 03722 South Korea}
    \IEEEauthorblockA{E-mail: yun-feng.lo@gatech.edu; \{cm.lee, cbchae\}@yonsei.ac.kr}
}

%% file: abstract.tex
\begin{abstract}
Molecular communication (MC), one of the emerging techniques in the field of communication, is entering a new phase following several decades of foundational research. Recently, attention has shifted toward MC in liquid media, particularly within tubular environments, due to novel application scenarios. The spatial constraints of such environments make accurate modeling of molecular movement in tubes more challenging than in traditional free-space channels. In this paper, we propose a three-dimensional channel model for molecular communications with an absorbing ring-shaped receiver in a tubular environment. To the best of our knowledge, this is the first theoretical study to model the impact of an absorbing ring-shaped receiver on the channel response in tube-based MC systems. The problem is formulated as a partial differential equation with heterogeneous boundary conditions, and an approximate solution is derived under flow-dominated conditions. The accuracy of the proposed model is validated through particle-based simulations. We anticipate that the results of this study will contribute to the design of practical MC systems in real-world tubular environments.
\end{abstract}

%% file: section1.tex
Molecular communications (MC), an emerging subfield within the broader discipline of communications, was initially developed to replace radio frequency (RF) communications in novel application areas at the micro- and nano-scales. In recent years, macro-scale MC gained increasing research interest~\cite{Farsad2016}. RF communications suffer from significant performance degradation in fluid environments, making conventional RF system designs unsuitable for use in fluid pipelines~\cite{Guo2021}. In contrast, molecular communications use molecules as information carriers, drawing inspiration from biological signaling systems such as the nervous and hormonal systems. Notably, many of these biological systems operate in tube-like environments (e.g., axons in the nervous system and blood vessels in the hormonal system)\cite{Jamali2019}. As a result, MC has the potential to outperform RF communications in similar tube-like environments, such as fluid pipelines. Recent research efforts have also explored using MC for biological applications in 6G systems, including health monitoring systems inside the human body\cite{Lee2023}.

To fully harness the potential of MC in tube-like environments, it is crucial to develop accurate channel models tailored to such scenarios. Several researchers have investigated channel models for MC that account for the spatial constraints inherent in tube environments, revealing that the resulting channel characteristics differ significantly from those in free-space environments~\cite{turan_channel_2018, wicke_modeling_2018, doustali_diffusive_2019, Jamali2019, Zoofaghari2019, dinc_general_2019, Lo2019, Arjmandi2021, schafer_transfer_2021, huang_molecular_2023}. The channel response is influenced by factors including tube dimensions and cross-sectional shape, flow type and velocity, as well as the boundary conditions (BCs) at the tube wall. The survey in~\cite{Jamali2019} provides an overview of channel models incorporating these factors. 

Various studies have explored different combinations of these three key factors—tube geometry, flow profile, and BCs—in modeling MC in tube environments.
Spatial constraints in MC channels are often modeled by assuming specific geometric shapes for tube cross-sections, such as cylindrical or rectangular pipes~\cite{Jamali2019}. For flow velocity profiles, the most common assumption is uniform flow—i.e., constant drift~\cite{Jamali2019}. However, more realistic profiles such as the Poiseuille flow field~\cite{wicke_modeling_2018,dinc_general_2019,Lo2019,schafer_transfer_2021} or even non-Newtonian fluid flows~\cite{huang_molecular_2023} have also been studied. Boundary conditions~\cite{turan_channel_2018,Lo2019} at the tube wall can be used to model the presence or absence of blockage~\cite{doustali_diffusive_2019} or biological receptors~\cite{Zoofaghari2019,Arjmandi2021}, which significantly affect the channel response. 

For MC in tube environments, we use the term ``homogeneous BCs" to refer to BCs that are uniform across the entire tube wall; ``heterogeneous BCs" refers to the opposite case. Generally, the task of accurately modeling the MC channel is significantly harder under heterogenous BCs. Early studies of MC in tube environments with heterogeneous BCs, e.g., with a partially-covering receiver~\cite{turan_channel_2018} or partial blockage~\cite{doustali_diffusive_2019}, use simulations to obtain a numerical channel response. 

Analytical channel modeling for MC in tube environments~\cite{Zoofaghari2019,dinc_general_2019,Lo2019,Arjmandi2021,schafer_transfer_2021}, before this work, focuses exclusively on homogeneous BCs. In~\cite{Zoofaghari2019}, the authors proposed a channel model for MC in a biological cylindrical environment with uniform flow and a homogeneous BC, incorporating molecular self-degradation effects. The studies in~\cite{dinc_general_2019,Lo2019} introduced approximate channel models for MC in a cylindrical tube, considering the Poiseuille flow field along with various homogeneous BCs. In~\cite{Arjmandi2021}, a model incorporating homogeneous BC and active transport of molecules through the tube wall was proposed. The study in~\cite{schafer_transfer_2021} solves a similar channel modeling problem exactly under a homogeneous BC at the tube wall.

These previous studies on tube environments have enabled MC researchers to envision a wide range of scenarios for applying molecular communication in various tubular and vessel-like settings. However, an additional step is necessary to develop analytical channel models that are realistic for such applications: moving beyond the idealized assumption of passive receivers. Passive receivers—also referred to as transparent or sensing receivers—do not interact with information molecules, making them challenging to realize in practical systems. In contrast, absorbing receivers (see, e.g.~\cite{turan_channel_2018, GENC2018, huang_molecular_2023, yilmaz2014three,lee2024characterizing}), which remove molecules upon arrival, can be implemented using, e.g. ligand-binding receptors~\cite{Zoofaghari2019}.

On the other hand, one could argue that a \emph{ring-shaped} receiver is more compatible with the geometry of a tube environment. In~\cite{turan_transmitter_2018}, a ring-shaped sensing receiver is used for the task of transmitter localization. From an engineering perspective, it is also more practical to consider a receiver that is attached to the inner tube wall (similar to, e.g.~\cite{wicke_modeling_2018}), especially when restricting the analysis to non-mobile receivers—as is done in this paper and many previous studies~\cite{Zoofaghari2019, dinc_general_2019, Lo2019, Arjmandi2021, schafer_transfer_2021}. In contrast, point receivers located inside the tube would need to be mobile, as they are subject to the flow velocity profile. 

Therefore, we propose the use of an absorbing ring-shaped receiver that is attached to the inner tube wall in cylindrical fluid environments.
The core challenge in mathematically modeling the MC channel in this setting lies in solving partial differential equations (PDEs) with heterogeneous BCs, where different segments of the domain boundary exhibit distinct behaviors. The first study in MC to theoretically address such BCs in \emph{spherical} environments appeared recently in~\cite{li_channel_2024}, where the authors employed a boundary homogenization technique to obtain an approximate solution for the molecular concentration. To the best of our knowledge, our work is the first to theoretically address heterogeneous BCs in cylindrical environments, proposing a novel analytical technique to find approximate solution formulas in the flow-dominated regime.

The main contributions of this paper are three-fold:
\begin{itemize}
    \item We formulate the channel modeling problem for MC systems in a three-dimensional tube environment with an absorbing ring-shaped receiver.
    
    \item We derive approximate analytical expressions for molecular concentration, arrival probability, and arrival rate in the flow-dominated regime.
    
    \item We perform particle-based simulations to validate the accuracy of the proposed analytical solutions.
\end{itemize}

%% file: section2.tex
We consider an MC system inside a straight cylindrical tube filled with a fluid medium. A transmitter (Tx) and a receiver (Rx) are fixed in position within the tube, which has a circular cross-section. \tc{black}{Following the molecular injection scenario considered in~\cite{Wicke2022}, it is assumed that an initial distribution of molecules is given along the cross-section at the Tx location, which does not affect the fluid flow.} For ease of analysis, \tc{black}{we assume that the initial distribution is concentrated at the center of the cross-section}; however, our channel modeling framework applies to Txs capable of generating an arbitrary initial concentration profile on a tube cross-section in an impulsive manner.

The Tx is located at the start of the tube, which we define as the cross-section at $z = 0$. For simplicity, we assume the tube is infinitely long, thereby neglecting fluid dynamical effects at the tube ends.  The Rx is modeled as a ring-shaped surface attached to the inner wall of the tube. The tube wall acts as a fully reflecting boundary for information molecules, except on the ring-shaped Rx region, which serves as a fully absorbing boundary to model molecular reception. The Rx spans from cross-section $z = d_1$ to $z = d_2$, and molecules absorbed by this surface are considered to be removed from the system.

In alignment with practical scenarios, we assume the fluid inside the tube exhibits steady, laminar flow with a constant axial velocity. The Tx is placed upstream relative to the Rx to ensure efficient transmission. The motion of each information molecule is modeled as an independent Brownian motion with constant drift. In the axial ($z$) direction, molecular movement is dominated by flow, whereas in the radial and angular directions, molecules undergo diffusion only.

\begin{figure}
	{\includegraphics[width=1\columnwidth,keepaspectratio]{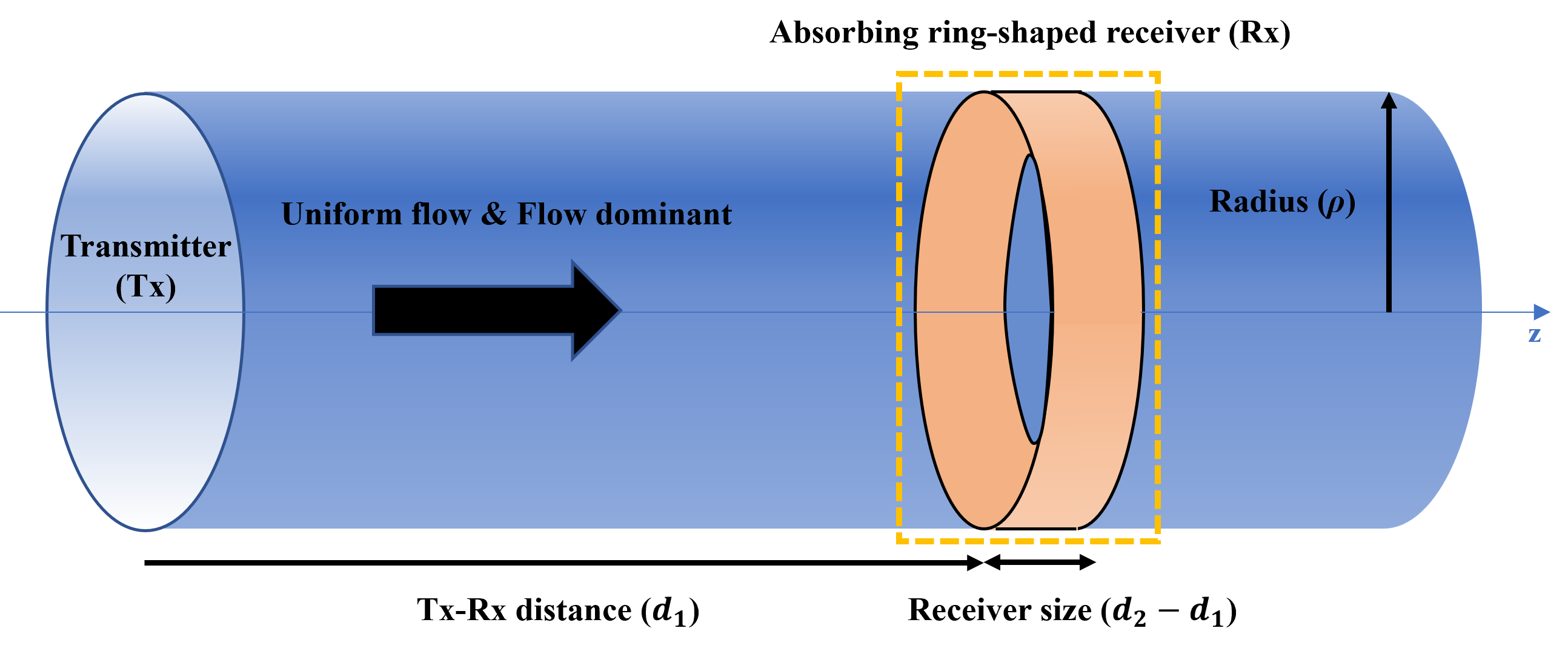}}
	\centering
	\caption{System model of MC in a tube with an absorbing ring-shaped receiver attached to the inner wall of the tube.}
	\label{Fig:system_model}
    \vspace{-10pt}
\end{figure}

To summarize, as illustrated in Fig.~\ref{Fig:system_model}, we consider an MC system within a cylindrical tube exhibiting uniform axial flow. Information molecules propagate under the combined effects of flow and diffusion. In the axial direction, flow is assumed to dominate diffusion, while molecules are reflected by the tube wall except at the absorbing ring-shaped Rx.

%% file: section3.tex
In this section, we present a general channel modeling framework applicable to MC systems in tubes with heterogeneous BCs in the flow-dominated regime.

%% file: section3a.tex
It is well known in the MC literature  (see, e.g.~\cite[Eq.~(17)]{Jamali2019}) that the molecular concentration $c=c(\boldsymbol{x}|t)$ at location $\boldsymbol{x}$ and time $t\geq 0$ satisfies the advection-diffusion PDE
\begin{equation}
\label{eq0}
\frac{\partial}{\partial t} c = D \nabla^2 c - \nabla \cdot (\boldsymbol{v} c)
\end{equation}
assuming a constant diffusion coefficient $D$ and a general flow velocity field $\boldsymbol{v}$, where $\frac{\partial}{\partial a}$ denotes partial differentiation with respect to (w.r.t.) variable $a$, $\nabla^2$ is the Laplacian operator and $\nabla\cdot$ is the divergence operator. Choosing the cylindrical coordinate system $\boldsymbol{x}=(r,\theta,z)$, the constant flow field is $\boldsymbol{v}=v\hat{\boldsymbol{z}}$ for some $v>0$, where $\hat{\boldsymbol{z}}$ is the unit vector in the positive axial direction. Then, \eqref{eq0} becomes the PDE
\begin{equation}
\label{eq1}
\frac{\partial}{\partial t} c = D \left(
\frac{\partial^2 }{\partial r^2} + 
	\frac{1}{r}\frac{\partial }{\partial \theta} +
	\frac{1}{r^2}\frac{\partial^2 }{\partial \theta^2} +
	\frac{\partial^2 }{\partial z^2}
    \right) c 
    - v \frac{\partial}{\partial z} c
,
\end{equation}
where $r\in[0,\rho]$, $\theta\in[0,2\pi)$, $z\in(-\infty,\infty)$, and $t>0$, with $\rho>0$ denoting the tube radius. 

\tc{black}{At time $t=0$, the initial distribution of molecules is assumed to be concentrated at the center of the cross-section, with a total of $N_\textnormal{e}$ molecules.} Thus the initial condition (IC) is
\begin{equation}
    \label{eq2}
    c(r,\theta,z|t=0) = N_\textnormal{e}
    \frac{\delta(r)}{2\pi r} \delta(z)
    .
\end{equation}

According to our system modeling with the absorbing ring-shaped receiver, the boundary conditions (BCs) are:
\begin{equation}
    \label{eq3}
    \frac{\partial c}  {\partial r}(r=\rho,\theta,z|t) = 0
    ,~\textnormal{for}~ z \in (-\infty,d_1) ,
\end{equation}
\begin{equation}
    \label{eq4}
    c(r=\rho,\theta,z|t) = 0 
    ,~\textnormal{for}~ z \in [d_1,d_2) ,
\end{equation}
\begin{equation}
    \label{eq5}
    \frac{\partial c}  {\partial r}(r=\rho,\theta,z|t) = 0 
    ,~\textnormal{for}~ z \in [d_2,\infty) ,
\end{equation}
\begin{equation}
    \label{eq6}
    \lim_{z\to\pm\infty} c(r,\theta,z|t) = 0
    = \lim_{z\to\pm\infty} \frac{\partial c}  {\partial z}(r,\theta,z|t) 
    ,
\end{equation}
where \eqref{eq6} specifies decay conditions along the tube axis. 

The proposed PDE formulation for molecular concentration is thus the PDE problem defined by the PDE~\eqref{eq1}, IC~\eqref{eq2} and BCs~\eqref{eq3}-\eqref{eq6}. To the best of our knowledge, no exact closed-form solution to this PDE problem is known. Specifically, due to the heterogeneous BCs in the axial direction, the separation of variables technique used in, e.g., \cite{Zoofaghari2019, Lo2019}, cannot be directly applied. On the other hand, by linearity of the PDE problem, we may assume $N_\text{e}=1$ without loss of generality.

We note that the proposed channel modeling framework can be extended to more general scenarios, including cases where each of the boundary conditions in~\eqref{eq3}--\eqref{eq5} is replaced with a Robin boundary condition~\cite{Zoofaghari2019,dinc_general_2019,Lo2019,Arjmandi2021}, and where the tube comprises more than three sections with heterogeneous boundary conditions.

%% file: section3b.tex
Assume that the system parameters are such that in the axial direction, the effect of flow for molecular movement dominates that of diffusion; we call this the ``flow-dominated regime." In this regime, it is reasonable to assume that the molecule passes through the cross-sections $z=d_1$ and $z=d_2$ only once, respectively (resp.); let $t_1$ and $t_2$, resp., denote the time instant at which the crossing happens. In the flow-dominated regime, we have that $t_1\approx d_1/v$ and $t_2\approx d_2/v$. 

Let us divide the time horizon $[0,\infty)$ into 3 periods: $[0,t_1)$, $[t_1,t_2)$, and $[t_2,\infty)$. We apply the heuristic that during each period, the molecular concentration can be approximated by a separation-of-variable type formula with
\begin{align}
    \label{eq:ansatz}
    c(r,\theta,z|t)
    =
    c_{r,\theta}(r,\theta|t) c_z(z|t)
    .
\end{align}
Applying~\eqref{eq:ansatz} in~\eqref{eq1}, we get that $c_{r,\theta}$ satisfies the PDE
\begin{equation}
    \label{eq10}
\frac{\partial}{\partial t} c_{r,\theta} = D \left(
\frac{\partial^2 }{\partial r^2} + 
	\frac{1}{r}\frac{\partial }{\partial \theta} +
	\frac{1}{r^2}\frac{\partial^2 }{\partial \theta^2}
    \right) c_{r,\theta}
.
\end{equation}
The IC~\eqref{eq2} (with $N_\textnormal{e}=1$) results in
\begin{equation}
    \label{eq14}
    c_{r,\theta}(r,\theta|t=0) = \frac{\delta(r)}{2\pi r}
    .
\end{equation}
Moreover, our heuristic asserts that the molecule ``effectively experiences" the corresponding BCs of that period. Specifically, the BC~\eqref{eq3} and~\eqref{eq5} translates to
\begin{align}
    \frac{\partial c_{r,\theta}}{\partial r}(r=\rho,\theta|t) = 0
    &~\textnormal{for}~t\in [0,t_1) \cup [t_2,\infty)
    \label{eq31}
,\end{align}
and the BC~\eqref{eq4} translates to
\begin{align}
    c_{r,\theta}(r=\rho,\theta|t) = 0
    &~\textnormal{for}~t\in [t_1,t_2)
    \label{eq29}
.\end{align}
 Finally, we piece the solutions for $c_{r,\theta}(r,\theta|t)$ in the three periods together by imposing the continuity of molecular concentration around $t=t_1$ and $t=t_2$. To be specific, we sequentially solve for $c_{r,\theta}$ in each period by applying the separation of variables method used in, e.g.~\cite{Zoofaghari2019}.

Due to space limitations, we omit details of the derivations. Let $J_\nu$ denote the order-$\nu$ Bessel function of the first kind~\cite{luke1962integrals}. Let
$j_{\nu,k}$ stand for the $k$-th $(k\geq 1)$ positive root of $J_\nu$ in ascending order, with $j_{\nu,0}:=0$. We write $j'_{\nu,k}=j_{\nu,k}/\rho$.

The approximate solution for $c_{r,\theta}$ in $t\in[0,t_1)$ is
\begin{equation}
    \label{eq32}
c_{r,\theta}(r,\theta|t\in [0,t_1)) 
= 
\sum_{n=0}^\infty 
\alpha_n
J_0\left( j'_{1,n} r \right) 
\textnormal{e}^{ -j'^2_{1,n} D t }
,
\end{equation}
where the coefficients $\alpha_n$, determined by the IC~\eqref{eq14}, are
\begin{equation}
    \label{eq38}
    \alpha_n = \left( \pi \rho^2\right)^{-1} \left( J_0(j_{1,n}) \right)^{-2}
\end{equation}
for $n\geq 0$.
The approximate solution for $c_{r,\theta}$ in $t\in[t_1,t_2)$ is
\begin{equation}
    \label{eq40}
c_{r,\theta}(r,\theta|t\in [t_1,t_2)) 
=
\sum_{m=1}^\infty 
\beta_m 
J_0\left( j'_{0,m} r \right) 
\textnormal{e}^{ -j'^2_{0,m} D (t-t_1) } 
.
\end{equation}
where $\beta_m$, determined by taking $\lim_{t\to t_1^-} c_{r,\theta}(r,\theta|t\in [0,t_1))$ as the IC for~\eqref{eq40}, are
\begin{align}
    \label{eq44}
    \beta_{m} 
    =
    \frac{2}{\pi \rho^2} 
    \frac{j_{0,m}}{J_1(j_{0,m})} 
    \sum_{n=0}^\infty 
    \frac{ 1 }{ J_0(j_{1,n}) }
    \frac{ \text{e}^{ - j'^2_{1,n} D t_1 } }{ j_{0,m}^2 - j_{1,n}^2 }
\end{align}
for $m\geq 1,$
where some integral identities for Bessel functions (see, e.g.~\cite{luke1962integrals}) are applied for simplification.
The approximate solution for $c_{r,\theta}$ in $t\in[t_2,\infty)$ is
\begin{align}
    \label{eq46}
	c_{r,\theta}(r,\theta|t\in [t_2,\infty)) 
	=
    \sum_{\ell=0}^\infty 
    \gamma_\ell 
    J_0\left(j'_{1,\ell} r\right) 
    \text{e}^{ - j'^2_{1,\ell} D (t-t_2) }
    .
\end{align}
Similarly, the coefficients $\gamma_\ell$ for $\ell\geq 0$ are determined by taking $\lim_{t\to t_2^-} c_{r,\theta}(r,\theta|t\in [t_1,t_2))$ as the IC for~\eqref{eq46}. After some integral identities, we get a simplified form for $\gamma_\ell$ as
\begin{align}
    \label{eq49}
    &
    \gamma_\ell 
    \nonumber\\
    =& 
    \frac{4}{\pi \rho^2} 
    \frac{1}{J_0(j_{1,\ell})} 
    \sum_{m=1}^\infty 
    \sum_{n=0}^\infty
    \frac{ j_{0,m}^2 }{ J_0(j_{1,n}) }
    \frac{ \text{e}^{ - j'^2_{0,m} D (t_2-t_1)} }{j_{0,m}^2-j_{1,\ell}^2} 
    \frac{ \text{e}^{ - j'^2_{1,n} D t_1 } }{j_{0,m}^2-j_{1,n}^2} 
    .
\end{align}

To obtain $c(r,\theta,z|t)$ approximately, we observe that in the flow-dominated regime, we have
\begin{align}
    \label{eq17}
    c_z(z|t) = \frac{1}{\sqrt{4\pi Dt}} \exp \left( - \frac{(z-vt)^2}{4Dt} \right)
    ,
\end{align}
which can be verified to satisfy the PDE for $c_z$ resulting from applying~\eqref{eq:ansatz} into~\eqref{eq1}, and the IC for $c_z$ at $t=0$ obtained from IC~\eqref{eq2} (with $N_\textnormal{e}=1$). Therefore, an approximate solution for molecular concentration (with a general $N_\text{e} \geq 1$) to the PDE problem \eqref{eq1}-\eqref{eq6} can be given as
\begin{align}
    \label{eq:approx_concen}
    c(r,\theta,z|t) 
    \approx
    N_\textnormal{e} 
    c_{r,\theta}(r,\theta|t) 
    c_z(z|t)
    .
\end{align}

%% file: section3c.tex
\begin{table}[t]
	\begin{center}
		\caption{Range of parameters used in the simulations}
		\renewcommand{\arraystretch}{1.14}
		\label{tbl:system_parameters}
		\begin{tabular}{p{4cm} l}
			\hline
			\bfseries{Parameter} 							& \bfseries{Value} \\ 
			\hline 
			Number of emitted molecules			& $1\,000$\\
			Duration of channel simulation
            & $\SI{3.5}{\second}$\\ 
			Time step of simulation & ${10^{-5}}\,\si{\second}$\\ 
			Replication 						& $100$\\
			Tx-Rx Distance ($d_1$) 				& $\{1\,000,\, 2\,000,\, 3\,000\}\,\, \si{\micro\metre} $\\
			Receiver size ($d_2 - d_1$) 				& $20\,\, \si{\micro\metre} $\\
			Tube length ($L$)		& $3\,500\,\,\si{\micro\metre}$\\
			Tube radius ($\rho$)			& $\{10,\, 20\}\,\,\si{\micro\metre}$\\
			Flow velocity ($v$) 	& $\{1\,000,\, 2\,000,\, 3\,000\}\,\,\si{\micro\metre/\second}$\\
			Diffusion coefficients ($D$) 	& $\{100,\, 400,\, 700\}\,\,\si{\micro\metre^2/\second}$\\
			\hline
		\end{tabular} 
	\end{center}
	\renewcommand{\arraystretch}{1}
	\vspace{-14pt}
\end{table}

We reinterpret the concentration $c(r,\theta,z|t)$ (with $N_\textnormal{e}=1$) in Subsec.~\ref{subsec:appr} as the conditional probability density, denoted $p_{R,\Theta,Z|T_1,T_2}(r,\theta,z|t;t_1,t_2)$, of random molecular position $(R,\Theta,Z)$ at time $t$ given the event $\{T_1=t_1,T_2=t_2\},$ where $T_1$ (resp. $T_2$) is the first arrival time of a molecule, released at the Tx at time $t=0$, to the cross-section $\{z=d_1\}$ (resp. $\{z=d_2\}$). Define $T$ as the first arrival time of the same molecule to the Rx; if the molecule never arrives, we set $T=\infty$. Then, for any $t\in [0,\infty)$ and $0\leq t_1\leq t_2<\infty$, define the conditional survival function $S(t|t_1,t_2):=\textnormal{Pr}\{T>t~|~T_1=t_1,T_2=t_2\}$ and the survival function $S(t):=\textnormal{Pr}\{T>t\}.$ A similar \emph{survival analysis} has been applied in the MC literature in, e.g.~\cite{srinivas2012molecular}. We have
\begin{equation}
    \begin{split}
    \label{eq53}
    &
    S(t|t_1,t_2)
    \\
    =~&
    \int_{-\infty}^\infty 
    \int_0^{2\pi} 
    \int_0^\rho 
    p_{R,\Theta,Z|T_1,T_2}(r,\theta,z|t;t_1,t_2) 
    ~r
    ~\mathrm{d}r
    ~\mathrm{d}\theta
    ~\mathrm{d}z  
    ,
    \end{split}
\end{equation}
and
\begin{equation}
    \label{eq54}
    S(t)
    = 
    \int_0^\infty 
    \int_{t_1}^\infty
    S(t|t_1,t_2) 
    f_{T_2|T_1}(t_2|t_1) 
    f_{T_1}(t_1) 
    ~\mathrm{d}t_2 
    ~\mathrm{d}t_1
    ,
\end{equation}
where $f_{T_1}$ is the PDF of $T_1$ and $f_{T_2|T_1}$ is the conditional PDF of $T_2$ given $T_1$. We know that $T_1\sim\textnormal{IG}[d_1/v,d_1^2/(2D)]$, where $\textnormal{IG}[\mu,\lambda]$ denotes the inverse Gaussian distribution with parameters $\mu$ and $\lambda$~\cite{srinivas2012molecular}. Moreover, defining $\Delta:=T_2-T_1$, strong Markov property (see, e.g.~\cite[Ch.2]{morters2010brownian}) of the axial Brownian motion (note that $\textnormal{Pr}\{T_1<\infty\}=1$) implies that $\Delta|\{T_1=t_1\}\sim\text{IG}[(d_2-d_1)/v,(d_2-d_1)^2/(2D)]$, the PDF of which we denote by $f_\Delta$ for short. Note that the expression for $f_{T_1}$ and $f_\Delta$ is well-known~\cite{srinivas2012molecular}, so we do not display them.

\begin{figure*}[t]
    \centering
    \scalebox{0.95}{
    \begin{minipage}{0.48\linewidth}
        \centering
        \includegraphics[width=\linewidth]{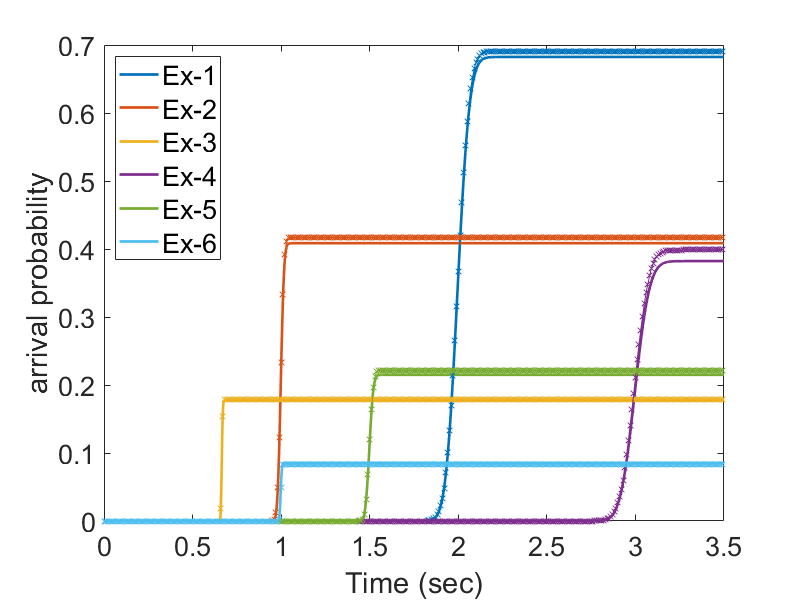}
        \subcaption{Arrival probability (theory) and absorption ratio (simulation).}
        \label{fig:result-cdf}
    \end{minipage}
    \begin{minipage}{0.48\linewidth}
        \centering
        \includegraphics[width=\linewidth]{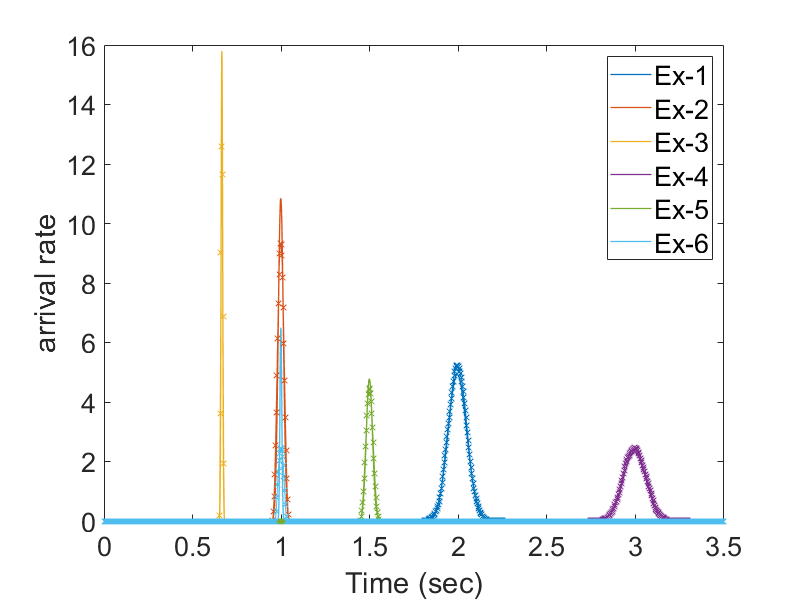}
        \subcaption{Arrival rate (theory) and instantaneous number of absorption (simulation).}
        \label{fig:result-pdf}
    \end{minipage}
    }
    \caption{Comparison of theoretical channel models and particle-based simulation results. “Ex-number” in the legend refers to the index of the parameter combinations listed in Table~\ref{tbl:system_parameters_2}.}
    \label{fig:result}
\end{figure*}

Using~\eqref{eq:approx_concen} (setting $N_\textnormal{e}=1$, with~\eqref{eq32}-\eqref{eq17}) as an approximation for $p_{R,\Theta,Z|T_1,T_2}(r,\theta,z|t;t_1,t_2)$ in~\eqref{eq53} and simplifying (details omitted), we obtain
\begin{equation}
    \label{eq76}
    S(t|t_1,t_2)
    =
    \sum_{m=1}^\infty
    \sum_{n=0}^\infty
        c_{m,n}
        \textnormal{e}^{
        -
        D
        \left( 
        j'^2_{1,n} t_1 
        + j'^2_{0,m}
        \int_0^t
        \mathbb{I}_{[t_1,t_2]}(\tau)
        \mathrm{d}\tau
        \right) 
        }
        ,
\end{equation}
where we denote the constants
\begin{align}
    \label{eq79}
    c_{m,n}
    &:=
    \frac{
    4
    }{
    \left(j_{0,m}^2 - j_{1,n}^2\right)J_0(j_{1,n})
    }
    &\text{for}~
    m\geq 1
    ,
    n\geq 0
    ,
\end{align}
and the indicator function
$\mathbb{I}_{\mathcal{S}}(\tau)=1$ if $\tau\in\mathcal{S}$ and zero otherwise, for $\mathcal{S}\subset[0,\infty)$. Applying~\eqref{eq76} into~\eqref{eq54}, we get
\begin{align}
    \label{eq77}
    &S(t)
    =
    \sum_{m=1}^\infty
    \sum_{n=0}^\infty
    c_{m,n}
    \int_0^\infty
    f_{T_1}(t_1)
    ~\textnormal{e}^{
        -
        D 
        j'^2_{1,n} t_1 
        }
    \int_0^\infty
    f_{\Delta}(\delta) 
    \nonumber\\
    &
    \qquad\qquad\qquad\times
        \textnormal{e}^{
        -
        D
        j'^2_{0,m}
        \int_0^t
        \mathbb{I}_{[t_1,t_1+\delta]}(\tau)
        \mathrm{d}\tau    
        }
        ~\mathrm{d}\delta ~\mathrm{d}t_1
        .
\end{align}
We define the \emph{arrival probability} $R(t):=1-S(t)=\textnormal{Pr}\{T\leq t\},$ i.e., the probability that each molecule arrives at the Rx before or at time $t\in[0,\infty)$. Intuitively, there is a nonzero probability that the molecule never arrives, due to axial flow dominance, implying $\lim_{t\to\infty} R(t)<1$. This intuition is consistent with our numerical evaluations in Fig.~\ref{fig:result-cdf} in Section~\ref{sec:resu}.

\begin{table}[t]
	\begin{center}
		\caption{Parameters for channel responses}
		\renewcommand{\arraystretch}{1.13}
            \label{tbl:system_parameters_2}
            \resizebox{\columnwidth}{!}{
		\begin{tabular}{c || c c c c | c c || c}
			\hline
            Ex&
            $\rho~(\si{\micro\metre})$ &
            $v~(\si{\micro\metre/\second})$ & 
            $D~(\si{\micro\metre^2/\second})$ &
            $d_1~(\si{\micro\metre})$   &
            $\mathrm{Re}$ &
            $\mathrm{Pe}$ &
            NRMSE
			\\
            \hline 
			1    &
            $10$    &
            $1,000$ &
            $700$   &
            $2000$  &
            $0.01$  &
            $14.2857$ &
            $0.9824$
            \\
			2    & 
            $10$    &
            $2,000$ &
            $400$   &
            $2000$  &
            $0.02$  &
            $50$    &
            $0.9610$
            \\
			3    &  
            $10$    &
            $3,000$ &
            $100$   &
            $2000$  &
            $0.03$  &
            $300$   &
            $0.9848$
            \\
            4    &  
            $20$    &
            $1,000$ &
            $700$   &
            $3000$  &
            $0.02$  &
            $28.5714$   &
            $0.9488$
            \\	
            5    &  
            $20$    &
            $2,000$ &
            $400$   &
            $3000$  &
            $0.04$  &
            $100$   &
            $0.9518$
            \\	
            6    &  
            $20$    &
            $3,000$ &
            $100$   &
            $3000$  &
            $0.06$  &
            $600$  &
            $0.9600$
            \\	
			\hline
		\end{tabular}
        }
	\end{center}
	\renewcommand{\arraystretch}{1}
	\vspace{-14pt}
\end{table}

\begin{table}[t]
	\begin{center}
		\caption{Metrics}
		\renewcommand{\arraystretch}{1.15}
            \label{tbl:metrics}
		\begin{tabular}{c || c c c c}
			\hline
            Metrics &
            Min &
            Max & 
            Mean &
            Std
			\\
            \hline 
			RMSE    &
            $0.0004$    &
            $0.0192$ &
            $0.0058$   &
            $0.0044$
            \\
			NMSE    & 
            $0.9708$    &
            $1.0000$ &
            $0.9966$   &
            $0.0049$
            \\
			NRMSE    &  
            $0.8293$    &
            $0.9959$ &
            $0.9504$   &
            $0.0315$
            \\
			\hline
		\end{tabular} 
	\end{center}
	\renewcommand{\arraystretch}{1}
	\vspace{-16pt}
\end{table}

Another quantity of interest is the \emph{arrival rate} of molecules to the Rx, which we denote by $r(t)=R'(t)$. Using Leibniz's integral rule and simplifying (details omitted), we obtain
\begin{align}
        \label{eq84}
        r(t)
        &=
        D
        \sum_{m=1}^\infty
        j'^2_{0,m}
        \textnormal{e}^{
        -
        D
        j'^2_{0,m} 
        t
        }
        \sum_{n=0}^\infty
        c_{m,n}
        \int_0^t
        f_{T_1}(t_1)
        \nonumber\\
        &\qquad\times
        \left(
        \int_{t-t_1}^\infty
        f_{\Delta}(\delta) 
        ~\mathrm{d}\delta 
        \right)
        \textnormal{e}^{
        -
        D
        \left( 
        j'^2_{1,n} 
        - j'^2_{0,m}
        \right)
        t_1
        }
        ~\mathrm{d}t_1
        .
\end{align}
It is known that he cumulative distribution function of inverse Gaussian distributions has a closed-form expression in terms of $Q$-functions $Q(x):=\int_x^\infty \textnormal{e}^{-\xi^2/2}~\mathrm{d}\xi/ \sqrt{2\pi}$ for $x\in\mathbb{R}$, i.e.,
\begin{align}
    \begin{split}
    \label{eq87}
    &
    \int_{t-t_1}^\infty 
    f_{\Delta}(\delta) ~\mathrm{d}\delta
    =
    Q\left(\frac{v(t-t_1)-(d_2-d_1)}{\sqrt{2D(t-t_1)}}\right)
    \\
    &~
    -
    \text{e}^
    {
    \frac{v(d_2-d_1)}{D}
    }
    Q\left(\frac{v(t-t_1)+(d_2-d_1)}{\sqrt{2D(t-t_1)}}\right)
    .
    \end{split}
\end{align}
Hence, for each $(m,n)$, the corresponding term in~\eqref{eq84} only requires a one-dimensional numerical integration to evaluate, assuming access to $Q$-function values.
Depending on how the Rx operates, either the arrival probability $R(t)$ or the arrival rate $r(t)$ may be considered the \emph{channel impulse response} of the proposed MC system.

%% file: section4.tex
\subsection{Parameters}
Our key assumptions of the theoretical channel model are uniform laminar flow and flow-dominant tube environments. We utilized particle-based simulation to confirm the theoretical channel model and we applied simulation parameters in Table~\ref{tbl:system_parameters}. Each parameter describes $\mu m$-scale and flow-dominant tube environments with uniform flow. We used Reynolds number and Péclet number to evaluate parameters that satisfied the above assumptions or not.

The Reynolds number is used to determine whether the flow is laminar or turbulent. The number is given as
\begin{equation}
	\mathrm{Re} = 
    \rho v/\nu, 
\end{equation}
where $\nu$ is the kinematic viscosity of the fluid which is roughly $10^{6}~\mu m^2/s$ in case of water~\cite{Wicke2022}. Generally, the fluid shows laminar flow when the Reynolds number is smaller than 2000, and the parameters in Table~\ref{tbl:system_parameters} satisfy the criterion~\cite{Jamali2019, Wicke2022}.

The Péclet number describes how dominant flow is over diffusion. The number is given as
\begin{equation}
	\mathrm{Pe} = 
    \rho v/D. 
\end{equation}
The parameters in Table~\ref{tbl:system_parameters} are set as appearing big Péclet number to satisfy flow-dominant assumption~\cite{Jamali2019}.

\subsection{Comparison}

We evaluated the particle-based simulation and the theoretical channel model using the parameters summarized in Table~I. The comparison was conducted using three performance metrics: root mean square error (RMSE), normalized mean square error (NMSE), and normalized RMSE (NRMSE). These metrics are defined as
\begin{equation}
	\textnormal{RMSE} = \sqrt{\frac{1}{N_\textnormal{S}} \sum_{i=1}^{N_\textnormal{S}} (x_i - y_i)^2}, 
\end{equation}
\begin{equation}
	\textnormal{NMSE} = 1 - \frac{\sum_{i=1}^{N_\textnormal{S}} (x_i - y_i)^2}{\sum_{i=1}^{N_\textnormal{S}} (y_i - \bar{y})^2},
\end{equation}
\begin{equation}
    \textnormal{NRMSE} = 1 - \sqrt{\frac{\sum_{i=1}^{N_\textnormal{S}} (x_i - y_i)^2}{\sum_{i=1}^{N_\textnormal{S}} (y_i - \bar{y})^2}},
\end{equation}
where $N_\textnormal{S}$ is the sample size, $x_i$ are the test data, $y_i$ are the reference data, and $\bar{y}$ is the mean value of the reference data~\cite{GENC2018}. We use these three metrics for measuring the overlap between the empirical arrival probability (reference data) and the theoretical arrival probability (test data). For RMSE, values closer to zero indicate a better fit. For NMSE and NRMSE, values closer to one indicates a better fit.

Due to the stochastic nature of particle-based simulations arising from Brownian motion, multiple trials (specifically, $100$, i.e., the Replication parameter in Table~\ref{tbl:system_parameters}) were conducted, and the results were averaged to reduce randomness. Meanwhile, as shown in~\eqref{eq77} and~\eqref{eq84}, the theoretical channel model contains infinite summation terms indexed by $n$ and $m$; numerically, it is only possible to compute an approximating finite sum, namely, summation up to resp. $N$ and $M$. We observed that increasing these values enhances the precision of the theoretical model in the sense of improved agreement with the simulation results. An example with $N=M=10$ is provided in Table~\ref{tbl:metrics}, demonstrating that the proposed theoretical model exhibits good consistency across the three metrics. The least accurate case was observed when the absorption ratio (or arrival probability) at the receiver was as low as $0.0634$, indicating minimal molecular absorption. This suggests that in scenarios with low expected absorption, larger $N$ and $M$ values are necessary to ensure accurate modeling.

Fig.~\ref{fig:result} presents six channel response examples using parameter sets defined in Table~\ref{tbl:system_parameters_2}. The theoretical channel model was evaluated with $N=M=10$, consistent with the setup in Table~\ref{tbl:metrics}. The simulation results represent the average of repeated trials. ``Ex-number" in graphs refers to the index of the parameter combinations in Table~\ref{tbl:system_parameters_2}. Theoretical responses are shown as solid lines, while simulation results are marked with crosses. Fig.~\ref{fig:result-cdf} illustrates the cumulative fraction of absorbed molecules over time, namely, the arrival probability, and Fig.~\ref{fig:result-pdf} shows the instantaneous number of absorbed molecules per time step, i.e., the arrival rate. As supported by the quantitative results in Table~\ref{tbl:metrics}, the theoretical model exhibits close agreement with particle-based simulations.

%% file: section5.tex

In this paper, we investigated the three-dimensional channel characteristics of molecular communication systems in a tubular environment featuring an absorbing ring-shaped receiver on the inner surface. A theoretical channel model was derived for the flow-dominated regime using a novel framework that combines the solution of partial differential equations with heterogeneous boundary conditions and survival analysis. The accuracy of the proposed model was validated through particle-based simulations. Our analysis focused on flow-dominated scenarios; extending the framework to diffusion-dominated regimes remains an open challenge. Nonetheless, the proposed model and derivation methodology provide valuable insights for the design of MC systems in confined tubular structures. We further envision that molecular communication in such environments will play a key role in the biological layer of 6G and beyond, enabling applications such as health monitoring via nano-machines implanted in human blood vessels.